\documentclass[fleqn,twoside]{article}
\usepackage[headings]{espcrc2}

\readRCS
$Id: espcrc2.tex,v 1.2 2004/02/24 11:22:11 spepping Exp $
\usepackage{graphicx, balance}
\usepackage[figuresright]{rotating}

\newcommand{\AmS}{{\protect\the\textfont2
  A\kern-.1667em\lower.5ex\hbox{M}\kern-.125emS}}

\title{\textbf{A Technique to Share Multiple Secret Images}}

\author{Mohit Rajput\address[DCSE]{Department of Computer Science and Engineering, National Institute of Technology, Uttarakhand, India-246174.\\},
Maroti Deshmukh\address{Assistant Professor, Department of Computer Science and Engineering, National Institute of Technology, Uttarakhand, India-246174.}}

\runtitle{A Technique to Share Multiple Secret Images}
\runauthor{Mohit Rajput, Maroti Deshmukh}

\usepackage{algorithm}
\usepackage{algpseudocode} 
\usepackage{subfigure}
\usepackage{cleveref}
\usepackage{amsmath}
\usepackage{algorithm}
\usepackage{url}
\usepackage{amsmath}
\usepackage{amssymb}
\usepackage{graphicx}

\begin{document}
\begin{abstract}

Visual Cryptography comes under cryptography domain. It deals with encrypting and decrypting of visual information like pictures, texts, videos $ etc.$ Multi Secret Image Sharing (MSIS) scheme is a part of visual cryptography that provides a protected method to transmit more than one secret images over a communication channel. Conventionally, transmission of a single secret image is possible over a channel at a time. But as technology grows, there emerge a need for sharing more than one secret image. An $(n, n)$-MSIS scheme is used to encrypt $n$ secret images into $n$ meaningless noisy images that are stored over different servers. To recover $n$ secret images all $n$ noisy images are required. At earlier time, the main problem with secret sharing schemes was that attacker can partially figure out secret images, even by getting access of $n-1$ or fewer noisy images. To tackle with this security issue, there arises a need of secure MSIS scheme, so that attacker can not retrieve any information by using less than $n-1$ noisy images. In this paper, we propose a secure $(n, n+1)$-MSIS scheme using additive modulo operation for grayscale and colored images. For checking the effectiveness of proposed scheme; Correlation, MSE and PSNR techniques are used. The experimental results show that the proposed scheme is highly secured and altering of noisy images will not reveal any partial information about secret images. The proposed $(n, n+1)$-MSIS scheme outperforms the existing MSIS schemes in terms of security. \\\\
{\bf Keywords :} Multi Secret Image Sharing (MSIS) Scheme,
Additive Modulo,  Floor, Ceil, Round, Correlation, RMSE,  PSNR. 

\end{abstract}

\maketitle

\section{INTRODUCTION}

In present era, with enhancement of technology usage, digital media also increases swiftly. This increase concern over security in digital media. Due to this concern various techniques for data hiding were introduced. Data hiding not only provide security  in various day to day applications like digital access for bank accounts, user authentication where high security is required, but also used in securing very sensitive and highly classified data like missile access codes, codes require to secure lockers etc.  Some of the data hiding techniques are Cryptography, Watermarking and Steganography. These methods are well known and highly used to hide the secret messages. Cryptography refers to process of converting plain text into encrypted form which is called as cipher text. In Cryptography we use keys to encrypt or decrypt data. Key refers to string of characters, which is used to decrypt or encrypt data at sender as well as receiver side. Main disadvantage associated with this method is sharing a key between sender and receiver. If some intruder gets access to the key, he can easily decode any secure message transfer between sender and receiver. Watermarking uses noise-tolerant signal and embeds it into digital media. The main disadvantage is, due to the introduction of noise-tolerant signal as it sometimes lead to error in decryption at receiver side. Steganography is an act to conceal secret data into other data. In this data hiding technique, the user hides the secret data into another data, known as a cover image. The disadvantage associated, is that unaltered recovery of data at the receiver side is very hard to achieve in stenography.
\newline
Visual Cryptography is a subdomain of Cryptography. Visual Cryptography is a technique of hiding visual data in such a way that when the correct shared images are stacked together it reveals the hidden visual data. Visual Cryptography, first makes its entry when Adi \cite{adi79} and Shamir \cite{naor95visual} proposed a method, where a secret image is encrypted into noisy images which do not reveal any information about secret images. Earlier, only single secret sharing schemes like $(n,n)$ and $(k,n)$ are used.  In $(n, n)$ single secret sharing schemes, the secret image is encrypted into $n$ shares and all the $n$ shares are required to recover the secret image, similarly in $(k, n)$  single secret sharing scheme, the secret image is encrypted into $n$ shares and at least $k$ shares are needed to recover the secret image, while less than $k$ shares are insufficient to decrypt the secret image. As the technology rises, their rises a demand for sharing more than one secret image at a time. For achieving this goal, multi secret sharing schemes like $(n,n)$ and $(n,n+1)$ were introduced. In $(n, n)$ multi secret sharing schemes, the $n$ secret images are encrypted into $n$ shares and all $n$ shares are required to recover the secret images, similarly in $(n, n + 1)$ multi secret sharing schemes, the $n$ secret images are encrypted into $n + 1$ shares and all $n + 1$ shares are required to recover the secret images. Questions may arise, why do we need another secure method for security when we have enough of them? How secret sharing schemes have advantages over others? If somehow any intruder gets access to some noisy images it can't fabricate secret image from them which can be easily done in case of cryptography. On receiver side, it can be easily reconstructed without loss or with negligible loss. Secret sharing scheme has many application fields, including missile launch codes, areas where trust plays an essential role, sharing data over untrusted channels, highly classified information, access control etc. To achieve higher reliability and confidentiality, we use secret sharing scheme as by storing noisy images on different data servers increases reliability as well as confidentiality. Rest of the paper structure is as follows. Section 2, discuss the previous work made in the area of secret sharing schemes. The proposed $(n, n+1)$-MSIS schemes are presented in Section 3. In Section 4, the experimental results and analysis are shown. Section 5 concludes the paper.

\section{Related Work}

Concept of visual cryptography, introduced by Shamir \cite{adi79} in $1979$ , proposed a method to share a secret using $(k, n)$-threshold scheme with $n = 2k - 1$. Blakley’s scheme \cite{blakley79} uses the concept of $n$-dimensional space to encrypt secret data in contrast to Shamir proposed method, which describe the secret data as the $y$-intercept of an $n$-degree polynomial. Both the secret sharing schemes uses the concept of $(k, n)$- secret sharing scheme. In $(k, n)$ secret sharing scheme, $n$ shared images are distributed among $n$ shareholders in such a manner that on combining $k$ shared images, it will reveal the secret. But, if less than $k$  shared images are combined then no secret is revealed. Chen et al.\cite{chen11} proposed $(n, n + 1)$-MSIS scheme based on simple Boolean XOR operation. In this scheme, $n$ secret images are used to create $n + 1$ shared images and to decode them, all $n + 1$ shared images are needed. In this scheme sharing capacity of multiple secret images are increased but it failed to produce randomized shared images because of simple Boolean XOR operation on secret images. Chen et al.\cite{chen14secure} presented a secure Boolean based $(n, n)$-MSIS scheme. In this scheme to increase the randomness in shared images bit shift function is used. This scheme requires more time because of bit shift function. Teng Guo et al.\cite{tzung2009} proposed a $(n, n)$ extended visual cryptography scheme. A secret image and $n$ cover images are encoded in $n$ share images in such a way that the stacking of all $n$ share images will reveal the secret image while from any less than $n$ share images, no information can be revealed. Lin et al.\cite{lin2014} proposed a novel random grid based MSIS scheme. Secrets images are encoded into two pie shared images and it can be decoded by stacking one pie share on another at different angle of rotation. Daoshun et al.\cite{daoshun2007} proposed $(n, n)$ scheme using XOR operation for gray scale images. In this $(n, n)$-MSIS scheme, $n$ secret images are encrypted into $n$ noisy images. No noisy image individually reveal any information about secret images but, if less than $n$ images are stacked over each other, partial information is revealed. Shyong et al.\cite{shyong2007} proposed a $(n, n)$- MSIS scheme using random grids for encryption of gray images as well as color images. Noisy images do not reveal any information when taken individually, whereas the secrets can be revealed when two noisy images are stacked over each other. Both of proposed method are accurate as no pixel expansion is seen. A $(k, n)$-RG based VSS scheme was proposed by Chen and Tsao et al.\cite{tzung2011} for binary and color images. A secret image is encrypted into $n$ meaning less random grids. This scheme uses $k$ shares to reveal secret image. Deshmukh et al.\cite{maroti2014} presents a comparative study of $(k, n)$ visual secret sharing scheme for binary images. Chen and Wu et al. presented a secure scheme by using bit shift function. Bit shift is used to generate random image to provide the randomness in noisy images. Deshmukh et al.\cite{maroti2016,maroticvip2016,marotiiciss2016} proposed a (n, n)-MSIS Scheme using boolean XOR and modular arithmetic operation. Additive modulo is used to reduce the computation time of the proposed method.

\section{Proposed Method}

With enhancement of technology, many MSIS scheme came into picture, some of them are $(n,n+1)$ and some are $(n,n)$ ~\textit{i.e.} they shares $n$ secret images among $n$ or $n+1$ receivers and to reconstruct these n secret images all $n$ or $n+1$ noisy images are required. An $(n,n+1)$-MSIS scheme is an n-out-of-(n+1) scheme. The main problem with many of the MSIS scheme was they reveal partial information from less than $n$ or $n+1$ noisy images, which compromises security. Chen et al.~\cite{chen14secure} –MSIS scheme reveal partial secret information from $(n-1)$ or fewer noisy images. Proposed scheme uses $n+1$ noisy shares to conceal $n$ secret images and no partial information can be retrieved from $n$ or less than $n+1$ noisy images. Proposed scheme uses additive modulo rather than XOR which is conventionally used. The main advantage of additive modulo over XOR operation is that it takes less time to execute. Boolean XOR operation has more time complexity than additive modulo as XOR perform bit by bit operation and XOR also reveals partial or complete information of secret image with less than $n$ noisy images. As we move towards color image from binary image, number of bits increase from \textit{1} to \textit{24}. We can easily figure out how rapidly time increases with increase in number of secret images and number of pixels in secret images. In additive modulo, there exists a unique additive inverse for every other element in the given range. Modular arithmetic are of two types; first is additive inverse and second is multiplicative inverse. In additive inverse, addition and modulo operations are used and in multiplicative inverse, multiplication and modulo operations are used. We say two numbers are additive inverse of each other if  $ a + b\equiv 0(mod\ n)$ where $b$ and $a$ are additive inverse of each other. Each integer has a unique additive inverse. For grayscale images and color images, pixel value ranges from $0-255$ and each number from $0-255$ has an additive inverse and its modulus value is $256$, whereas each number may or may not have a multiplicative inverse in this range. We have used additive inverse rather than multiplicative inverse. 
                        
In this proposed scheme, $n$ secret images $SI_i$, $i=1,2,\cdots,n$ are encrypted into $n+1$ noisy images $NI_i$, $i=1,2,\cdots,n+1$. Firstly, Temporary shares $C_i$, $i=1,2,\cdots,n$ are created by performing division operation on secret images $SI_i$, $i=1,2,\cdots,n$ with divisor as $ n+1 $. To truncate floating points into respective closest integers we use round function, round function as it takes closest integer value and provide more precise results than ceil or floor function as shown in Table~\ref{tab:one}. Division operation is performed to reduce the pixel values. This is done so that, pixel values do not exceed from $255$, when some scalar constant ($a \in N$) is multiplied with them. In second step, a random matrix $R$ is created. In third step a server side key $SK$ is generated by using additive modulo operation on $C_i$, $i=1,2,\cdots,n$. In final step, noisy images $NI_i$, $i=1,2,\cdots,n+1$ are generated. $n$ noisy images are generated by using additive modulo operation on temporary shares which is generated in step one $C_i$, $i=1,2,\cdots,n$, server side key $SK$ and random matrix $R$ ~\textit{so,}. Last noisy image $NI_{n+1}$ is generated by using additive modulo operation on $n+1$ times of server side key $SK$ and $n+1$ times of random matrix. The main purpose of $(n+1){th}$ noisy image generation is to find random matrix $R$ at receiver side. The noisy image generation algorithm of proposed $(n, n+1)$-MSIS scheme is given in Algorithm ~\ref{alg:1}.

\begin{algorithm*}[htb]
        \caption{: \textbf{Proposed Noisy Share Generation Procedure.}}
        \vspace{1 mm}
        \textit{
        \textbf{Input:}   $n$ Secret images \{${SI_1,SI_2 \cdots SI_n}$\} of size $h \times w$. \\
        \textbf{Output:} $n+1$ Noisy images \{${NI_1,NI_2 \cdots NI_n,NI_{n+1}}$\}.
               } 
        \begin{algorithmic}
        \State  \textit{ 1. Generate $n$ Temporary Shares \{${C_1,C_2 \cdots C_n}$\} using round function.
                          \State	 \hspace{6mm}$C_i=round((SI_i/(n+1))) $ , 
                       where \{${i=1,2, \cdots, n}$\}}
                          
        \State  \textit{2. Generate a Random Matrix $R$ of size $h \times w$
        				  \State	\hspace{6mm}$R = Random(h,w)$}
        
        \State  \textit{3. Generate Server Side Key $SK$ using Additive Modulo
                          \State	 \hspace{6mm}$SK=(C_1) mod \ 256$ 
                          \State	 \hspace{6mm}$SK=(C_i + SK) mod \ 256$, where \{${i=2,3, \cdots, n}$\} }  
        \State  \textit{ 4. Generate $n+1$ Noisy images  \{${NS_1,NS_2 \cdots NS_n,NS_n+1}$\} using Additive Modulo
                          \State	\hspace{6mm}$NI_i= (C_i + SK + R)mod \ 256$	 where \{${i=1,2, \cdots, n}$\}   
                          \State	\hspace{6mm}$NI_{n+1}= ((n+1)\times (SK + R))mod \ 256$}
                          	
		\end{algorithmic}
\label{alg:1} 
\end{algorithm*}

The recovery procedure is different from encryption algorithm. This provides additional security as if, any intruder gets access to encryption algorithm then also he can't retrieve the secret information from noisy images. In recovery procedure we retrieve $n$ secret images from $n+1$ noisy images. In first step, generation of client side key $CK$ is done by performing additive inverse operation on first $n$ noisy images $NI_i$, $i=1,2,\cdots,n$. In second step we generate temporary noisy images $P_i$, $i=1,2,\cdots,n$ by performing multiplication on noisy images $NI_i$, $i=1,2,\cdots,n$ by $(n+1)$ and using modular operation, here $(n+1)$ is the number of noisy images send by sender. Third step deals with generation of random matrix $R$ used at server end to provide randomness in pattern for noisy images. Random matrix $R$ is generated by using additive inverse operation on $(n+1)^{th}$ noisy image $NI_{n+1}$ and client side key $CK$. Finally in fourth step, recovered images $RI_i$, $i=1,2,\cdots,n$ which is same as that of secret images. Recovered images $RI_i$, $i=1,2,\cdots,n$ are generated by using additive inverse operation on temporary noisy images $P_i$, $i=1,2,\cdots,n$, client side key $CK$ and random matrix $R$. The recovery procedure of proposed $(n, n+1)$-MSIS scheme is given in Algorithm~\ref{alg:2}.

\begin{algorithm*}[htb]
        \caption{: \textbf{Proposed Recovery Procedure.}}
        \vspace{1 mm}
        \textit{
        \textbf{Input:}  $n+1$ Noisy images \{${N_1,N_2 \cdots N_n}$\}. \\ 
        \textbf{Output:} $n$ Recovered images \{${RI_1,RI_2 \cdots RI_n}$\}. 
         }      
        \begin{algorithmic}
        
       \State \textit{  1. Genrate Client Side Key $CK$ 
                      \State    \hspace{6mm}$CK= (N_1) mod \ 256$
                      \State    \hspace{6mm}$CK= (CK + N_i) mod \ 256$, where \{${i=2,3 \cdots, n}$\}}
                         
       \State \textit{  2. Generate Temporary Noisy Images 
                      \State	 \hspace{6mm}$P_i=((n+1)\times N_i) mod \ 256$, where \{${i=1,2, \cdots, n}$\}}
               
       \State \textit{  3. Generation of Random matrix $R$ using Additive Inverse 
                      \State	 \hspace{6mm}$R=(N_{n+1} - CK) mod \ 256$    }                 
       \State \textit{  4. Generation of Recovered images\{ $RI_i$, $i=1,2,\cdots,n$\}
                      \State	 \hspace{6mm}$RI_i=(P_{i} - (CK + R))mod \ 256$  }

        \end{algorithmic}
        \label{alg:2} 
        \end{algorithm*}

\section{Experimental Results and Analysis}
In this Section, experimental results and analysis of proposed $(n, n+1)$-MSIS is done. The experiments are performed for grayscale and colored images. For binary images we have to make some changes in algorithm, modulus value should be updated as \textit{2} with it, multiplication and division operator has to be taken off. Experimental results are performed on Intel(R) Core(TM) i7-4710HQ 2.50Ghz Processor, 8GB RAM machine using MATLAB 13. All images are of size $512 \times 512$.

\begin{figure*}[!t]
\center
\subfigure[$SI_1$]{\includegraphics[width=3.5cm,height=4cm,keepaspectratio]{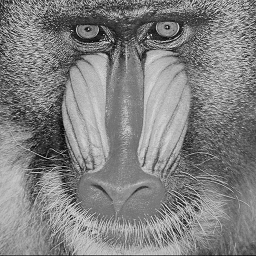}}
\hfill
\subfigure[$SI_2$]{\includegraphics[width=3.5cm,height=4cm,keepaspectratio]{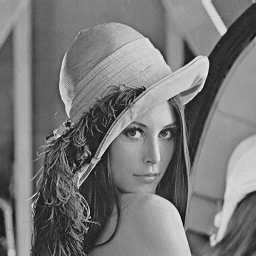}}
\hfill
\subfigure[$SI_3$]{\includegraphics[width=3.5cm,height=4cm,keepaspectratio]{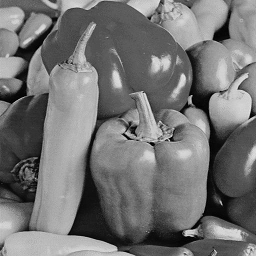}}
\hfill
\subfigure[$SI_4$]{\includegraphics[width=3.5cm,height=4cm,keepaspectratio]{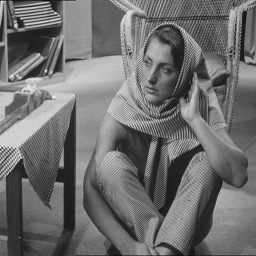}}
\hfill
\subfigure[$SI_5$]{\includegraphics[width=3.5cm,height=4cm,keepaspectratio]{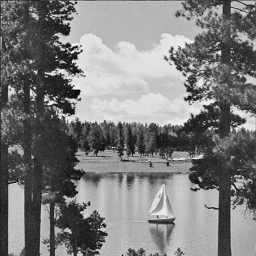}}
\hfill
\subfigure[$NI_1$]{\includegraphics[width=3.5cm,height=4cm,keepaspectratio]{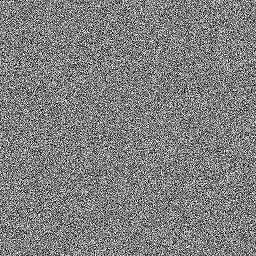}}
\hfill
\subfigure[$NI_2$]{\includegraphics[width=3.5cm,height=4cm,keepaspectratio]{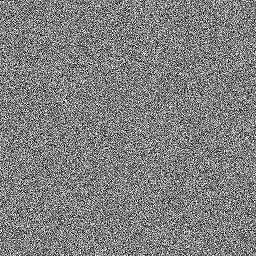}}
\hfill
\subfigure[$NI_3$]{\includegraphics[width=3.5cm,height=4cm,keepaspectratio]{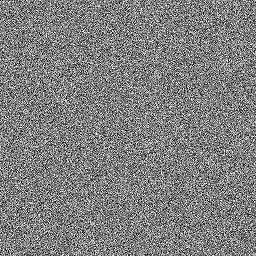}}
\hfill
\subfigure[$NI_4$]{\includegraphics[width=3.5cm,height=4cm,keepaspectratio]{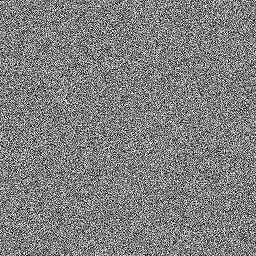}}
\hfill
\subfigure[$NI_5$]{\includegraphics[width=3.5cm,height=4cm,keepaspectratio]{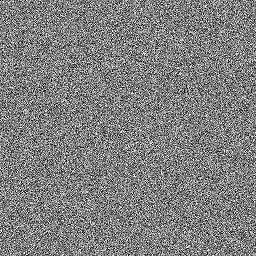}}
\hfill
\subfigure[$NI_6$]{\includegraphics[width=3.5cm,height=4cm,keepaspectratio]{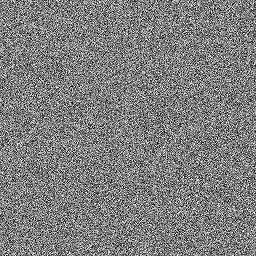}}
\hfill
\subfigure[$RI_1$]{\includegraphics[width=3.5cm,height=4cm,keepaspectratio]{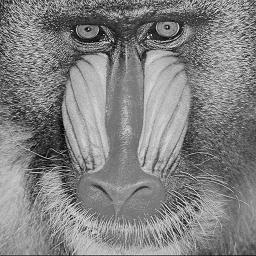}}
\hfill
\subfigure[$RI_2$]{\includegraphics[width=3.5cm,height=4cm,keepaspectratio]{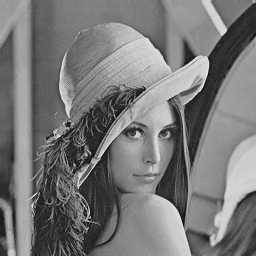}}
\hfill
\subfigure[$RI_3$]{\includegraphics[width=3.5cm,height=4cm,keepaspectratio]{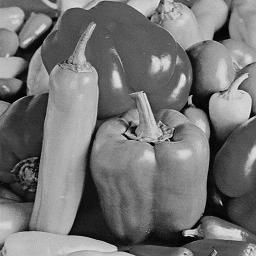}}
\hfill
\subfigure[$RI_4$]{\includegraphics[width=3.5cm,height=4cm,keepaspectratio]{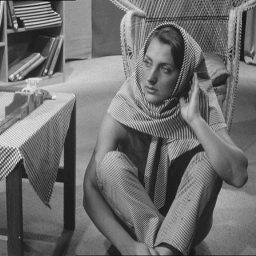}}
\hfill
\subfigure[$RI_5$]{\includegraphics[width=3.5cm,height=4cm,keepaspectratio]{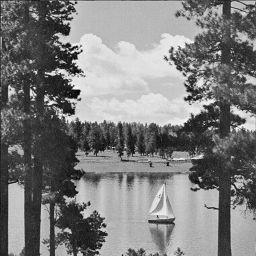}}
\hfill
\caption { Result of Proposed $(n, n+1)$-MSIS scheme for grayscale images with n=5: (a-e) Secret images ($SI_1, SI_2, SI_3, SI_4, SI_5$); (f-k) Noisy images ($NI_1, NI_2, NI_3, NI_4, NI_5, NI_6$); (l-p) Recovered images ($RI_1, RI_2, RI_3, RI_4, RI_5$).}
\label{fig:one}
\end{figure*}

The experimental results of proposed $(n, n+1)$-MSIS scheme for grayscale images are shown in Fig.~\ref{fig:one}. Input secrets images $SI_1, SI_2, SI_3, SI_4, SI_5$ are shown in Fig.~\ref{fig:one}(a)-(e) respectively. Fig.~\ref{fig:one}(f)-(k) shows noisy images $ NI_1, NI_2, NI_3, NI_4, NI_5, NI_6$ respectively. No share individually reveals any information of secret images. Fig.~\ref{fig:one}(l)-(p) shows recovered images $RI_1, RI_2, RI_3, RI_4, RI_5$ which are almost similar to the secret images.

\begin{figure*}[!t]
\center
\subfigure[$SI_1$]{\includegraphics[width=3.5cm,height=4cm,keepaspectratio]{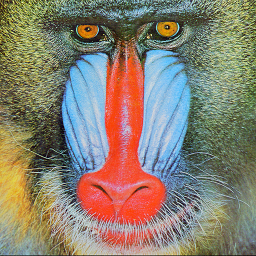}}
\hfill
\subfigure[$SI_2$]{\includegraphics[width=3.5cm,height=4cm,keepaspectratio]{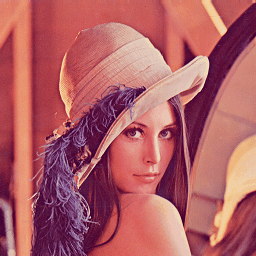}}
\hfill
\subfigure[$SI_3$]{\includegraphics[width=3.5cm,height=4cm,keepaspectratio]{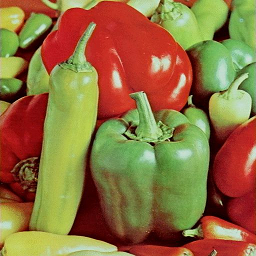}}
\hfill
\subfigure[$SI_4$]{\includegraphics[width=3.5cm,height=4cm,keepaspectratio]{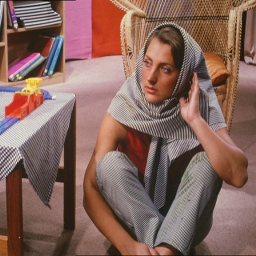}}
\hfill
\subfigure[$SI_5$]{\includegraphics[width=3.5cm,height=4cm,keepaspectratio]{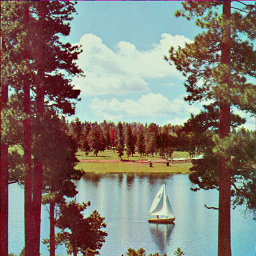}}
\hfill
\subfigure[$NI_1$]{\includegraphics[width=3.5cm,height=4cm,keepaspectratio]{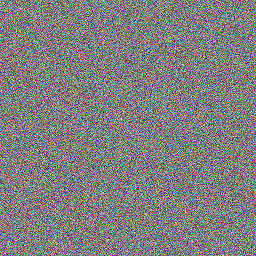}}
\hfill
\subfigure[$NI_2$]{\includegraphics[width=3.5cm,height=4cm,keepaspectratio]{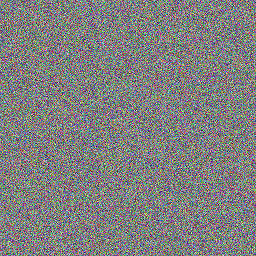}}
\hfill
\subfigure[$NI_3$]{\includegraphics[width=3.5cm,height=4cm,keepaspectratio]{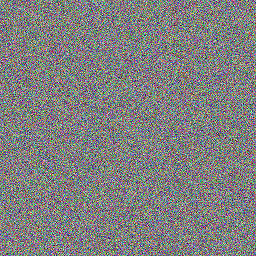}}
\hfill
\subfigure[$NI_4$]{\includegraphics[width=3.5cm,height=4cm,keepaspectratio]{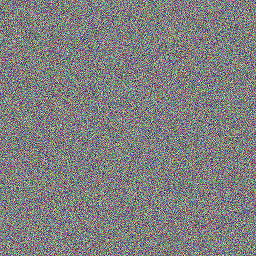}}
\hfill
\subfigure[$NI_5$]{\includegraphics[width=3.5cm,height=4cm,keepaspectratio]{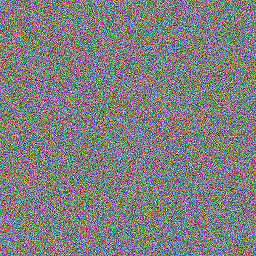}}
\hfill
\subfigure[$NI_6$]{\includegraphics[width=3.5cm,height=4cm,keepaspectratio]{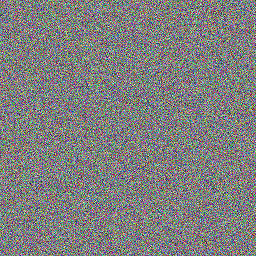}}
\hfill
\subfigure[$RI_1$]{\includegraphics[width=3.5cm,height=4cm,keepaspectratio]{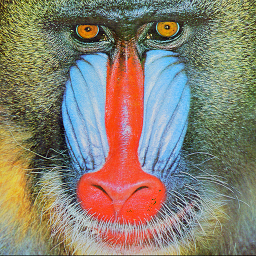}}
\hfill
\subfigure[$RI_2$]{\includegraphics[width=3.5cm,height=4cm,keepaspectratio]{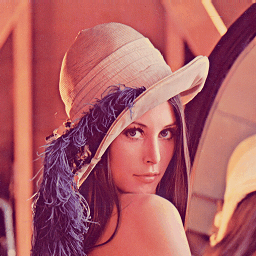}}
\hfill
\subfigure[$RI_3$]{\includegraphics[width=3.5cm,height=4cm,keepaspectratio]{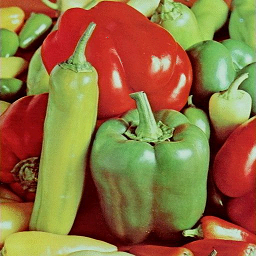}}
\hfill
\subfigure[$RI_4$]{\includegraphics[width=3.5cm,height=4cm,keepaspectratio]{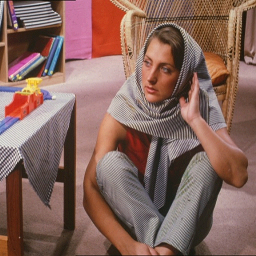}}
\hfill
\subfigure[$RI_5$]{\includegraphics[width=3.5cm,height=4cm,keepaspectratio]{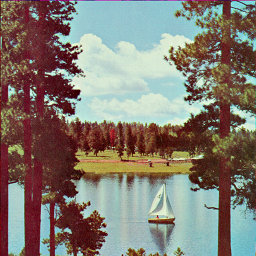}}
\hfill
\caption { Result of Proposed $(n, n+1)$-MSIS scheme for color images with n=5: (a-e) Secret images ($SI_1, SI_2, SI_3, SI_4, SI_5$); (f-k) Noisy images ($NI_1, NI_2, NI_3, NI_4, NI_5, NI_6$); (l-p) Recovered images ($RI_1, RI_2, RI_3, RI_4, RI_5$).}
\label{fig:two}
\end{figure*}

The experimental results of proposed $(n, n+1)$-MSIS scheme for colored images are shown in Fig.~\ref{fig:two}. Input secrets images $ SI_1, SI_2, SI_3, SI_4, SI_5 $ are shown in Fig.~\ref{fig:two}(a)-(e) respectively. Fig.~\ref{fig:two}(f)-(k) shows noisy images $ NI_1, NI_2, NI_3, NI_4,$ $NI_5, NI_6$ respectively. No share individually reveals any information of secret images. Fig.~\ref{fig:two}(l)-(p) show recovered images $RI_1, RI_2, RI_3, RI_4, RI_5$ and recovered images are almost similar to the secret images.

\subsection{Quantitative Analysis}

In quantitative analysis, similarity between secret and recovered images of proposed $(n, n+1)$-MSIS scheme is done using Correlation, RMSE and PSNR. The results we get using different functions like floor, ceil and round are shown in Table~\ref{tab:one}.\\

\textbf{Correlation:} The correlation coefficient $r$ value lies between $+1$ and $-1$. If $r$ is equals to $+1$, it indicates that the two compared images are same, if $r$ is equals to $-1$, it indicates that both the compared images are opposite to each other and $r$ equals to $0$ indicates that the compared images are uncorrelated. Correlation coefficient, $r$ is given as $:$
\tiny
\begin{equation}
 r = \frac{n \sum p q - (\sum p)(\sum q)}{\sqrt{(n \sum p^2-(\sum p)^2)(n \sum q^2-(\sum q)^2)  }}
\end{equation}
\normalsize
Here, $n =$ no of paired score; $\sum p q =$ sum of the product of paired score; $\sum p = $sum of $p$ scores; $\sum q = $sum of $q$ scores; $\sum p^2 = $sum of squared $p$ scores; $\sum q^2 = $sum of squared $q$ scores.\\

\textbf{RMSE:} RMSE is the root mean square error between the original image $I$ and the compared image $R$. RMSE is used as quality measure, ~\textit{i.e.} it is used to check the similarity between the two data sets. Low RMSE value correspond to greater similarity. RMSE is given as$:$
\tiny
\begin{equation}
RMSE=\sqrt{\frac{1}{M\times N}\sum_{x=1}^{M}\sum_{y=1}^{N}\left ( SI(x,y)-RI(x,y) \right )^{2}} 
\end{equation}
\normalsize
Where $M \times N$ is the dimension of image; $SI(x,y)$ pixel value at $(x,y)$ position for secret image; $RI(x, y)$  pixel value at $x,y$ position for recovered secret image.\\

\textbf{PSNR:} PSNR stands for peak signal to noise ratio. It is measured in decibel(dB). PSNR measures the quality of the recovered images and secret images. The higher the PSNR better the quality and vice versa. The PSNR is given as:
\tiny
\begin{equation}
PSNR(dB)=20 \ log_{10}\frac{255}{RMSE}
\end{equation} 
\normalsize
Where $RMSE$ is Root Mean Square Error. $255$ is used as we have used grayscale and colored images for analysis purpose.; $20$ decreases the RMSE difference between two images $10$ times.

\begin{table*}
\caption{Comparison of Secret and Recovered Images of Proposed $(n, n+1)$-MSIS scheme.}
\renewcommand{\baselinestretch}{1}
\resizebox{0.67\textwidth}{!}{\begin{minipage}{\textwidth}
\begin{tabular} {c c c c c c c c c c}
\hline
\centering \textbf{Secret and  Recovered Images} & \multicolumn{3}{c}{\textbf{FLOOR}}&\multicolumn{3}{c}{ \textbf{CEIL}} &\multicolumn{3}{c}{ \textbf{ROUND}}\\
\cline{2-10}
&  $Correlation$ &  $RMSE$ &  $PSNR$ &  $Correlation$ &  $RMSE$ &  $PSNR$  & $Correlation$ &  $RMSE$ &  $PSNR$  \\
\hline
 $SI_1, RI_1$ & $0.9989$ & $3.6069$ & $37.02dB$ & $0.9992$ & $3.0288$ & $38.54dB$  & 
$0.9992$ & $1.7796$ & $43.16dB$\\

 $SI_2, RI_2$ & $0.9994$  & $2.8873$ & $38.96dB$  & $0.9994$ & $3.1793$ & $38.12dB$ & $0.9994$ & $1.7796$  & $43.10dB$\\

 $SI_3, RI_3$ & $0.9995$  & $3.0325$ & $38.53dB$  & $0.9995$ & $3.0270$ & $38.54dB$ & $0.9995$ & $1.7808$  & $43.15dB$\\

$SI_4, RI_4$  & $0.9993$ & $3.0230$ & $38.56dB$ & $0.9993$ & $3.0306$ & $38.53dB$ & $0.9993$ & $1.7808$  & $43.15dB$\\

 $SI_5, RI_5$  & $0.9997$ & $3.0251$ & $38.55dB$  & $0.9997$ & $3.0293$ & $38.54dB$ & $0.9997$ & $1.7803$  & $43.15dB$\\

 $SI_6, RI_6$  & $0.9997$ & $3.0251$ & $38.55dB$  & $0.9997$ & $3.0293$ & $38.54dB$ & $0.9997$ & $1.7803$  & $43.15dB$\\
\hline
\end{tabular}
\end{minipage}}
\label{tab:one}
\end{table*}

Correlation is used as a quality measure to measure the similarity between secret images  $SI_i$, $i=1,2,\cdots,n$ and recovered images  $RI_i$, $i=1,2,\cdots,n$,  getting by stacking less than $n+1$ noisy images  $NI_i$, $i=1,2,\cdots,n$. Correlation coefficient $r$, for the following data set is approximately equals to $0$ , as shown in Table~\ref{tab:two} , this indicates that the secret images and recovered images are uncorrelated to each other. For less than $n+1$ noisy images, less than $n$ recovered images are generated. For Example, if we take 3 noisy images then only 2 recovered images will be generated. Due to this reason the underline quality measure can't be applicable, hence to represent the following case in Table~\ref{tab:two}, we have represented it with $N.A\hspace{1.5mm}(Not \hspace{1.5 mm} Applicable)$ state. 
    
\begin{table*}[!t]
\caption{Comparison of Secret Images with Recovered Images getting by different combination of Noisy Images.}
\resizebox{0.95\textwidth}{!}{\begin{minipage}{\textwidth}
\renewcommand{\baselinestretch}{1}
\centering
\begin{tabular} {l c c c c c}

\hline
\multicolumn{1}{l}{\textbf{Various Combination of Noisy Images}} & \multicolumn{5}{c}{\textbf{Secret and Noisy Image}}\\ 
\cline{2-6}
& $SI_1, RI_1$ &  $SI_2, RI_2$ &  $SI_3, RI_3$ &  $SI_4, RI_4$ &  $SI_5, RI_5$   \\
\hline
 $NI_1, NI_2, NI_3, NI_4, NI_5$ & $-0.0001$ & $-0.0002$ & $-0.0008$ & $0.0031$ & $0.0036$\\

 $NI_1, NI_2, NI_3, NI_4$ & $-0.0003$ & $-0.0019$ & $-0.0011$ & $0.0020$ & $N.A$\\

 $NI_1, NI_2, NI_3$ & $-0.0012$ & $-0.0004$ & $-0.0024$ & $N.A$ & $N.A$\\

 $NI_1, NI_2$  & $0.0032$ & $0.0014$ & $N.A$ & $N.A$ & $N.A$\\

 $NI_1$  & $-0.0011$ & $N.A$ & $N.A$ & $N.A$ & $N.A$\\
\hline

\end{tabular}
\label{tab:two}
\end{minipage}}
\end{table*}

A complete analysis regarding performance of proposed $(n, n+1)$-MSIS scheme is given in Table~\ref{tab:three}. No pixel expansion is there as size of secret images, noisy images and recovered images are same. To reveal secrets all $n+1$ shares are needed. For making $(n,n+1)$-MSIS scheme work for binary images, some changes are to be make, like modulus value has to be updated as $2$. No scalar multiplication and division is needed as it will lead to overflow of bits. Additive modulo is used for encryption and decryption. Sharing capacity of proposed scheme is $n/n+1$.

\begin{table}
\renewcommand{\baselinestretch}{1}
\caption{Analysis of proposed $(n, n+1)$-MSIS scheme.}
\begin{small}
\begin{center}
\begin{tabular}{l l} \hline
    $\textbf{Parameters}$ & $\textbf{Proposed Scheme Result}$ \\ \hline 
 $Secret Images  $   &   $n$  \\ 
 $Shared Images$   &   $n+1$ \\ 
 $Time(s)$   &   $0.329 \hspace*{0.1cm} Second$ \\ 
 $Pixel Expansion$   &   $No$ \\ 
 $Reveal Secrets$   &   $No$ \\ 
 $Sharing Type$   &   $Rectangle$  \\ 
 $Sharing Capacity$   &   $n/n+1$ \\ 
 $Color depth$   &   $Binary, Grayscale, Color$ \\ 
 $Recovery Stragey$   &   $Additive Modulo$ \\ \hline
 
\end{tabular}
\label{tab:three}
\end{center}
\end{small}
\end{table}

\begin{figure}
\centering
\includegraphics[width=2.5in]{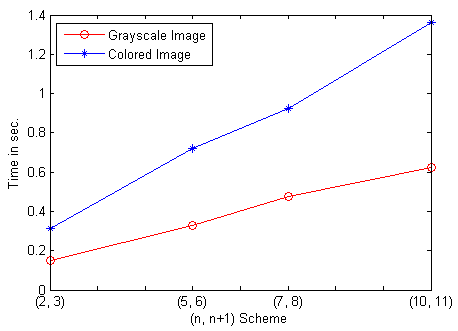}
\caption{Computation time of Proposed Scheme.}\label{fig:time}
\label{fig:three}
\end{figure}

Time complexity of proposed $(n, n+1)$-MSIS scheme for grayscale and colored secret images is shown in Fig.~\ref{fig:three}. As we increase no of secret images ~\textit{i.e. (value of n)} time required for execution also increase both for color and grayscale images. Computation time for colored image is more than binary and grayscale image because increase in number of bits. Time complexity of proposed scheme is high due to multiplication and division operations performed during encryption and decryption.

\section{Conclusion}
In this Paper, we overcome the security problem which was faced in ~\cite{chen11,chen14secure,yang2015} MSIS schemes. We used additive modulo operation which is faster than XOR operation. The proposed scheme shows better results in terms of security. Proposed scheme uses random matrix to generate randomness in shared images, so that stacking of less than $n+1$ noisy images will not reveal any information of secret images. To check the similarity between secret and recovered images we used Correlation, RMSE and PSNR techniques.

\newpage
\noindent{\includegraphics[width=1in,height=1.7in,clip,keepaspectratio]{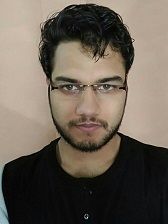}}
\begin{minipage}[b][1.9in][c]{4in}
{\centering{\bf {Mohit Rajput}} is an undergraduate student of Department of Computer Science and Engineering, National Institute of Technology, Uttarakhand.}\\\\
\end{minipage} \\\\
\noindent{\includegraphics[width=1in,height=1.7in,clip,keepaspectratio]{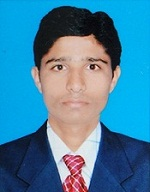}}
\begin{minipage}[b][1in][c]{4in}
{\centering{\bf{Maroti Deshmukh}} Maroti Deshmukh is
currently working as an Assistant Professor in Dept. of Computer Science and Engineering, National Institute of Technology, Uttarakhand. He obtained his B.Tech in CSE from SGGSIE\&T Nanaded. He received his M.Tech in IT from Hyderabad Central University, Hyderabad. His research area includes Secret Sharing Schemes, Cryptography, Image Security, Image Processing.}\\\\
\end{minipage}\\\\


\begin{thebibliography}{}
\bibitem{adi79}
Shamir, Adi. ``How to share a secret." Communications of the ACM 22.11 (1979): 612-613.

\bibitem{blakley79}
Blakley, George Robert. ``Safeguarding cryptographic keys.” Proc. of the National Computer Conference1979 48 (1979): 313-317.

\bibitem{naor95visual} 
Naor, Moni, and Adi Shamir. ``Visual cryptography." Advances in Cryptology—EUROCRYPT'94. Springer Berlin/Heidelberg, 1995.

\bibitem{chen11} 
Chen, Tzung-Her, and Chang-Sian Wu. ``Efficient multi-secret image sharing based on Boolean operations." Signal Processing 91.1 (2011): 90-97.
 
\bibitem{chen14secure} 
Chen, Chien-Chang, and Wei-Jie Wu. ``A secure Boolean-based multi-secret image sharing scheme." Journal of Systems and Software 92 (2014): 107-114.

\bibitem{yang2015} 
Yang, Ching-Nung, Cheng-Hua Chen, and Song-Ruei Cai. ``Enhanced Boolean-based multi secret image sharing scheme." Journal of Systems and Software (2015).

\bibitem{daoshun2007} 
Wang, Daoshun, et al. ``Two secret sharing schemes based on Boolean operations." Pattern Recognition 40.10 (2007): 2776-2785.

\bibitem{shyong2007} 
Shyu, Shyong Jian. ``Image encryption by random grids." Pattern Recognition 40.3 (2007): 1014-1031.

\bibitem{tzung2011}
Chen, Tzung-Her, and Kai-Hsiang Tsao. ``Threshold visual secret sharing by random grids." Journal of Systems and Software 84.7 (2011): 1197-1208.

\bibitem{maroti2014}
Maroti Deshmukh, Munaga V.N.K. Prasad. ``Comparative Study of Visual Secret Sharing Schemes to Protect Iris Image.” International Conference on Image and Signal Processing (ICISP), (2014): 91-98.

\bibitem{lin2014}
Lin, Kai-Siang, Chih-Hung Lin, and Tzung-Her Chen. ``Distortionless visual multi-secret sharing based on random grid." Information Sciences 288 (2014): 330-346.

\bibitem{tzung2009}
Tzung-Her Chen and Kai-Hsiang Tsao. ``Visual secret sharing by random grids revisited." Pattern Recognition, 42(9):2203–2217, 2009.

\bibitem{maroti2016}
 Maroti Deshmukh, Neeta Nain, and Mushtaq Ahmed. ``An (n, n)-Multi Secret Image Sharing Scheme Using Boolean XOR and Modular Arithmetic." 2016 IEEE 30th International Conference on Advanced Information Networking and Applications (AINA), pp.690-697, 2016.

\bibitem{maroticvip2016}
Maroti Deshmukh, Neeta Nain, and Mushtaq Ahmed. ``A Novel Approach of an (n, n) Multi Secret Image Sharing Scheme using Additive Modulo." International Conference on Computer Vision and Image Processing (CVIP), Springer, 2016.

\bibitem{marotiiciss2016}
Maroti Deshmukh, Neeta Nain, and Mushtaq Ahmed. ``Enhanced Modulo based Multi Secret Image Sharing Scheme." International Conference on Information Security Systems (ICISS), Springer, 2016.
\end{thebibliography}
\end{document}